\documentstyle[preprint,epsf,aps]{revtex}
\begin{document}
\title{ A theorem for the beam splitter entangler
}
\author{Wang Xiang-bin\thanks{email: wang$@$qci.jst.go.jp} 
\\
        Imai Quantum Computation and Information project, ERATO, Japan Sci. and Tech. Corp.\\
Daini Hongo White Bldg. 201, 5-28-3, Hongo, Bunkyo, Tokyo 113-0033, Japan}

\maketitle 
\begin{abstract}
It is conjectured that the an entanglement output states from a beam splitter requires
the non-classicality in the input state(M.S. Kim, W. Son, V. Buzek and P. L. Knight, Phys. Rev. A, 65, 032323(2002)). 
Here we give a proof for this conjecture. 
\end{abstract}
The beam splitter is one of the few quantum devices that may act as the entangler.
The entangler properties of a beam splitter have been studied in the past\cite{tan,sanders,paris,kim}. 
In particular, Kim et al\cite{kim} studied the entangler property with many different input states such as
the Fock number state, the coherent state, the squeezed state and the mixed states in Gaussian
form. It has been conjectured there that, to obtain an entangled output state,
 a necessary condition is that the input state should be non-classical. 
Unfortunately, there is no proof for this conjecture. In this paper, we give a very simple
proof for this conjecture. \\
Consider a loseless beam splitter(see figure 1). We can distinguish the field  mode $a$ and mode
$b$ by the different propagating direction. Most generally,  the property of a beam splitter
operator $\hat B$ in Schrodinger picture can be summarized by the following equations(see e.g., ref\cite{campos0})
\begin{eqnarray}
\rho_{out}=\hat B \rho_{in} \hat B^{-1},
\end{eqnarray} 
\begin{eqnarray}
\hat B^\dagger=\hat B^{-1},
\end{eqnarray}
\begin{eqnarray}
\hat B \left(\begin{array}{c}\hat a\\\hat b\end{array}\right)\hat B^{-1}
=M_{B}\left(\begin{array}{c}\hat a\\ \hat b\end{array}\right)\label{m1},
\end{eqnarray}
\begin{eqnarray}
M_B
=\left(\begin{array}{cc}\cos\theta e^{i\phi_0}& \sin\theta e^{i\phi_1}\\ -\sin\theta e^{-i\phi_1}
& \cos\theta e^{-i\phi_0} \end{array}\right)
\label{m2}\end{eqnarray} 
\begin{eqnarray}
\hat B |00\rangle=|00\rangle\label{va}.
\end{eqnarray}
Here $\rho_{in}$ and $\rho_{out}$ are the density operator for the input 
and output states respectively. Both of them are two mode states including mode $a$
and mode $b$. The elements in the matrix $M_B$ 
are determined by the beam splitter itself, 
$\hat a,\hat b$ are the annihilation operators for  mode $a$ and mode $b$
respective, $|00\rangle$ is the vacuum
state for both mode.  Equation (\ref{va}) is due to the simple fact of no input no output.
Without any loss of generality, we can express $\rho_{in}$
by the $P$ representation in the following form:
\begin{eqnarray}
\rho_{in}=\int_{-\infty}^{\infty} 
P(\alpha_a,\alpha_b,\alpha_a^*,\alpha_b^*) 
|\alpha_a,\alpha_b\rangle\langle\alpha_a,\alpha_b|{\rm d^2}\alpha_a
{\rm d^2} \alpha_b, 
\end{eqnarray}
where $|\alpha_a,\alpha_b\rangle$ is the coherent state in two mode Fock space, i.e.
\begin{eqnarray}
|\alpha_a,\alpha_b\rangle=\hat D_{ab}(\alpha_a,\alpha_b)|00\rangle 
\end{eqnarray}
and
\begin{eqnarray}
\hat D_{ab}(\alpha_a,\alpha_b)
=e^{\hat a^\dagger \alpha_a-\hat a \alpha_a^*+\hat b^\dagger \alpha_b-\hat b \alpha_b^*  }.
\end{eqnarray} 
If $\rho_{in}$ is a classical state, the distribution function $P(\alpha_a,\alpha_b,\alpha_a^*,\alpha_b^*) $
must be non-negative definite in the whole complex planes. In such a case, the ouput state is
\begin{eqnarray}
\rho_{out}= \int_{-\infty}^{\infty} 
P(\alpha_a,\alpha_b,\alpha_a^*,\alpha_b^*)\hat B
 |\alpha_a,\alpha_b\rangle\langle\alpha_a,\alpha_b|\hat B^{-1}{\rm d^2}\alpha_a
{\rm d^2} \alpha_b\end{eqnarray}
which is equivalent to
\begin{eqnarray}
\rho_{out}=\int_{-\infty}^{\infty} 
P(\alpha_a,\alpha_b,\alpha_a^*,\alpha_b^*)\hat B \hat D_{ab}(\alpha_a,\alpha_b) \hat B^{-1}\cdot \hat B
|00\rangle\langle 00|\hat B^{-1}\cdot \hat B \hat D_{ab} \hat B^{\dagger}. 
\end{eqnarray} 
From equation(\ref{va}) we know that $\hat B
|00\rangle\langle 00|\hat B^{-1}=|00\rangle\langle 00|$. By equation(\ref{m1}) we can see that
\begin{eqnarray}
\hat B \hat D_{ab}(\alpha_a,\alpha_b) \hat B^{-1}=\hat D_{ab}(\alpha'_a,\alpha'_b)
\end{eqnarray}
and
\begin{eqnarray}
(\alpha_a',\alpha_b')=(\alpha_a,\alpha_b)M_B.
\end{eqnarray}
In short, the following equatoin can be easily obtained by equation(\ref{m1},\ref{m2},\ref{va} )
\begin{eqnarray}
\hat B|\alpha_a,\alpha_b\rangle\langle\alpha_a,\alpha_b|\hat B^{-1}=
|\alpha_a',\alpha_b'\rangle\langle\alpha_a',\alpha_b'|
\end{eqnarray} 
Since ${\rm det} M_B=1 $,  we have the following formula for the output state
\begin{eqnarray}
\rho_{out}=\int_{-\infty}^{\infty} 
P(\alpha_a,\alpha_b,\alpha_a^*,\alpha_b^*) 
|\alpha_a',\alpha_b'\rangle\langle\alpha_a',\alpha_b'|{\rm d^2}\alpha_a'
{\rm d^2} \alpha_b'.
\end{eqnarray}
This is equivalent to
\begin{eqnarray}
\rho_{out}=\int_{-\infty}^{\infty} 
P'(\alpha_a,\alpha_b,\alpha_a^*,\alpha_b^*) 
|\alpha_a,\alpha_b\rangle\langle\alpha_a,\alpha_b|{\rm d^2}\alpha_a
{\rm d^2} \alpha_b\label{sep}
\end{eqnarray}
and 
\begin{eqnarray}
P'(\alpha_a,\alpha_b,\alpha_a^*,\alpha_b^*) 
=P(\alpha_a'',\alpha_b'',\alpha_a''^*,\alpha_b''^*) 
\end{eqnarray}
and
\begin{eqnarray}
(\alpha_a'',\alpha_b'')=(\alpha_a'',\alpha_b'')M_B^{-1}.
\end{eqnarray}
Since $P(\alpha_a,\alpha_b,\alpha_a^*,\alpha_b^*)$ is non-negative, 
functional $P'(\alpha_a,\alpha_b,\alpha_a^*,\alpha_b^*)$ must be also non-negative. 
By the definition of separability, state $\rho_{out}$ defined by equation(\ref{sep})
must be separable. Therefore we have the following theorem:\\
{\bf Theorem:} {\it If the input state is a classical state, the output state of a beam splitter must be 
a separable state.} \\
This is equivalent to say, in order to obtain an entangled output
state, the non-classicality of the input state is a necessary condition. This theorem can be extended
to a more general situation in the {\it multi}mode Fock space. Let's consider the rotation operator $\hat R$
in n-mode Fock space. We have  
\begin{eqnarray}
\hat R \Lambda  \hat R^{-1}=M_R \Lambda,
\end{eqnarray}
where $\Lambda=(\hat c_1,\hat c_2\cdots \hat c_n)^T$, $\hat c_i$ are the annihilation operators of the $i'$th mode and
$M_R$ is a $n-$dimensional unitary matrix. By using the BCH formula 
\begin{eqnarray}
e^\mu \nu e^{-\mu}=\nu + [\mu,\nu]+\frac{1}{2!}[\mu,[\mu,\nu]] +\cdots
\end{eqnarray}
we have the following explicitly formula for the operator $\hat R$:
\begin{eqnarray}
\hat R =\exp\left(-\Lambda^\dagger \ln M_R \Lambda\right).
\end{eqnarray}
Therefore we know that
\begin{eqnarray}
\hat R |00\cdots 0\rangle =|00\cdots 0\rangle.
\end{eqnarray}
Any classical multimode state in Fock space can be written in the following probabilistic
distribution
\begin{eqnarray}
\rho = \int_{-\infty}^{\infty}\bf P({\bf \alpha, \alpha^*}) |{\bf \alpha}\rangle\langle \bf \alpha |
\rm d^2\bf \alpha,
\end{eqnarray}
where $|\bf \alpha\rangle=|\alpha_1 \alpha_2\cdots\alpha_n\rangle$ and $\bf P({\bf \alpha, \alpha^*}) $
is a non-negative functional provided that $\rho$ is a classical state. 
Similar to the two mode case, we can  shown that, 
\begin{eqnarray}
\rho = \int_{-\infty}^{\infty}\bf P'({\bf \alpha, \alpha^*}) |{\bf \alpha}\rangle\langle \bf \alpha |
\rm d^2\bf \alpha,
\end{eqnarray}
and
\begin{eqnarray}
\bf P'({\bf \alpha, \alpha}^*) 
=\bf P({\bf \alpha}'', 
{\bf \alpha}^{''*}) 
\end{eqnarray}
and
\begin{eqnarray}
(\alpha'')=( \alpha)\cdot M_R^{-1}.
\end{eqnarray}
Obviously, the funtional $\bf P'({\bf \alpha, \alpha^*}) $ is non-negative.
Thus we draw the following conclusion in the multimode Fock space:
{\it  A classical density operator in the multimode Fock space
is separable under arbitrary rotation}.\\
 Although the nonclassicality in the input state
is a necessary condition, it is obviously not a sufficient condition for the entanglement
in the output state of a beam splitter. Since a beam splitter operater
 is  unitary, it is reversible. It has been shown in ref\cite{kim} that nonclassical
separable input state can be changed to an entangled state in the output. The inverse  
of such a process makes examples that even though the input state is nonclassical, the 
output could be still separable. Some specific examples are given in \cite{campos}.\\ 
I thank Prof Imai for support. I thank Dr Huang WY, Dr Winter, Dr Yura H, Dr Matsumoto K,and Dr Tomita A for useful
discussions.

\begin{figure}
\begin{center}
\epsffile{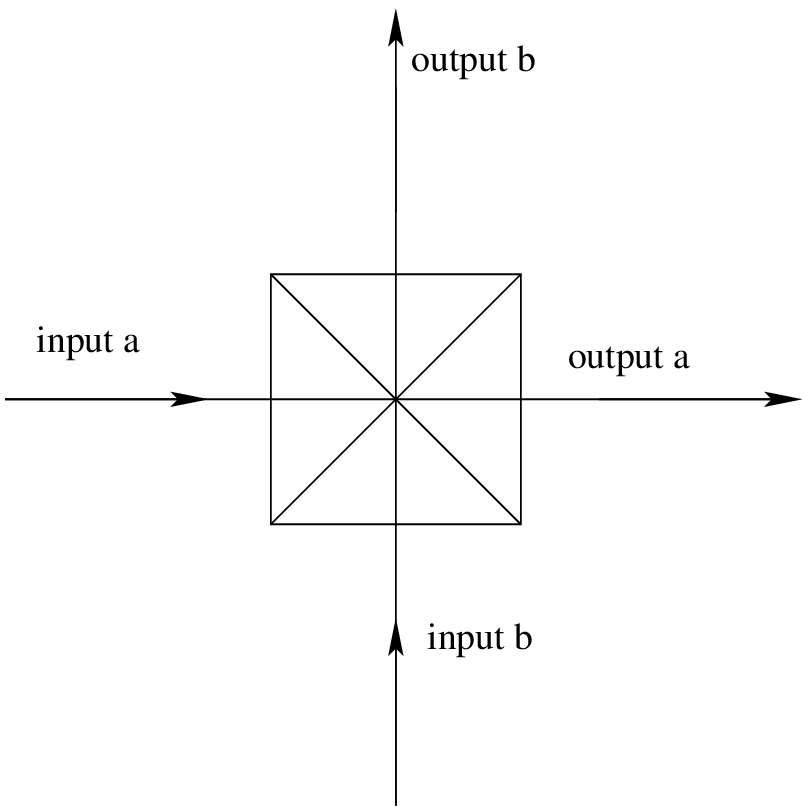}
\end{center}
\caption{ A schematic diagram for the beamsplitter operation. Both the input and the output
are two mode states. The different mode is distinguished by the propagating direction of the 
field.} 
\end{figure}

\end{document}